\documentstyle[aps,prl,floats]{revtex}

\newcommand{\bastar}{\begin{eqnarray*}}
\newcommand{\eastar}{\end{eqnarray*}}
\newskip\humongous \humongous=0pt plus 1000pt minus 1000pt

\newif\ifdtup

\relax
\newcommand{\be}{\begin{equation}}
\newcommand{\ee}{\end{equation}}
\newcommand{\bea}{\begin{eqnarray}}
\newcommand{\eea}{\end{eqnarray}}
\newcommand{\X}{{\vec X}}
\newcommand{\pro}{\partial}
\newcommand{\n}{\hat n}
\newcommand{\oneg}{\displaystyle\frac{1}{g}}

\newcommand{\D}{{\hat D}}

\newcommand{\A}{{\vec A}}
\newcommand{\valpha}{{\vec \alpha}}

\newcommand{\dfrac}{\displaystyle\frac}
\newcommand{\ba}{\begin{array}}
\newcommand{\ea}{\end{array}}

\newcommand{\nn}{\nonumber}
\begin{document}
\twocolumn[\hsize\textwidth\columnwidth\hsize\csname@twocolumnfalse%
\endcsname
\title  {Faddeev-Niemi Conjecture and Effective Action of QCD}
\bigskip

\author{Y. M. Cho$^{1}$, H. W. Lee$^{2}$, and D. G. Pak$^{3,4}$}

\address{
$^{1)}$Department of Physics, College of Natural Sciences, Seoul National University,
Seoul 151-742, Korea  \\
$^{2)}$Department of Physics, College of Natural Sciences, Chungbuk 
University, Chungju, Korea \\
$^{3)}$Asia Pacific Center for Theoretical Physics, 
207-43 Cheongryangri-dong, Dongdaemun-gu,
Seoul 130-012, Korea \\
$^{4)}$Department of Theoretical Physics, Tashkent State University, Tashkent
700-095, Uzbekistan \\
{\scriptsize \bf ymcho@yongmin.snu.ac.kr,
dmipak@apctp.kaist.ac.kr} \\ \vskip 0.3cm
}
\maketitle

\begin{abstract}
We calculate a one loop effective action of $SU(2)$ QCD in the presence of
the monopole background, and find a possible connection between   
the resulting QCD effective action and a generalized Skyrme-Faddeev action of
the non-linear sigma model.
The result is obtained using the gauge-independent decomposotion
of the gauge potential into the topological degrees
which describes the non-Abelian monopoles and the local dynamical degrees of
the potential, and integrating out all the dynamical degrees of QCD.

\vspace{0.3cm}
\end{abstract}

\narrowtext
\bigskip
                           ]
Recently Faddeev and Niemi have discovered the knot-like topological solitons 
in the Skyrme-type non-linear sigma model~\cite{faddeev1}, 
and made an interesting conjecture 
that the Skyrme-Faddeev action could be interpreted as an effective 
action for QCD in the low energy limit~\cite{faddeev2}. On the other hand 
it is generally believed that QCD in the low energy limit 
must exhibit the confinement of color, and it has long been argued 
that the confinement                       
could take place through the condensation of the monopole
which generates the dual Meissner effect~\cite{nambu,cho1}. 
So, in order to reconcile the Faddeev-Niemi conjecture with the confinement,  
one must be able to construct the effective action of QCD from the first 
principles and produce the mass scale that the Skyrme-Faddeev 
action contains, which could demonstrate the dynamical symmetry breaking  
and triggers the confinement of color in 
QCD \cite{cho1,cho2}. However, a rigorous derivation of 
an effective action of QCD which could 
generate the desired monopole condensation 
has been extremely difficult~\cite{savv,sch}. 
Fortunatly  
several authors have recently been able to argue that one can indeed
derive a generalized Skyrme-Faddeev action from the effective action of QCD in 
the infra-red limit, at least in some approximation~\cite{cho3,lang}.  

The purpose of this paper is to study the Faddeev-Niemi conjecture
in more detail. Our analysis shows that, after the monopole condensation,
the effective action of QCD could be approximated around the new vacuum
by a generalized Skyrme-Faddeev action.
Furthermore, our analysis strongly indicates
that the mass scale in the Skyrme-Faddeev action is closely related to
the confinement scale in QCD.  This establishes
a deep connection between a generalized
Skyrme-Faddeev type non-linear sigma model and QCD 
which is remarkable. On the other hand the assertion that the Skyrme-Faddeev
action could describe QCD in the infra-red limit is found to be misleading.
In particular, it is highly unlikely that the Faddeev-Niemi knots
can be realized in QCD.

To study a dynamical symmetry breaking induced by the monopole 
condensation one must first 
calculate the effective action of QCD
under the monopole background.
To do this one has to integrate
out all the dynamical degrees of the non-Abelian gauge theory,
leaving only the monopole configuration as the background.
For this it is necessary for us to identify the
monopole potential, and separate it from the generic
QCD connection, in a gauge independent manner. This can be done
with the gauge-independent decomposition of the non-Abelian 
gauge potential into the dual potential and the valence
potential \cite{cho1,cho2},
which provides us a natural reparameterization
of the non-Abelian connection in terms of
the dual potential of the maximal Abelian
subgroup $H$ of the gauge group $G$ and the gauge covariant vector
field of the remaining $G/H$ degrees. 
With this decomposition one can calculate the effective action of QCD
using the background field method, and show that
the monopole condensation takes place after one
integrates out all the dynamical degrees of the non-Abelian gauge
potential. 

For simplicity we will consider $SU(2)$ QCD.  A natural way to identify the
monopole potential and to separate the topological degree
is to introduce an isotriplet unit vector field
$\n$ which selects the color direction at each space-time point, and to
decompose the connection into the Abelian projection which leaves $\n$
invariant and the remaining part
which forms a covariant vector field \cite{cho1,cho2},
\bea
 & \vec{A}_\mu =A_\mu \n - \oneg \n\times\pro_\mu\n+\X_\mu\nonumber
         = \hat A_\mu + \X_\mu, \nn\\
 &  (A_\mu = \n\cdot \vec A_\mu, ~ \n^2 =1,~ \hat{n}\cdot\vec{X}_\mu=0),
\eea
where $A_\mu$
is the ``electric'' potential. 
Notice that the Abelian projection $\hat A_\mu$ is 
precisely the connection which
leaves $\n$ invariant under the parallel transport,
\bea
\D_\mu \n = \pro_\mu \n + g {\hat A}_\mu \times \n = 0.
\eea
Under the infinitesimal gauge transformation
\bea
\delta \n = - \vec \alpha \times \n  \,,\,\,\,\,
\delta \A_\mu = \oneg  D_\mu \vec \alpha,
\eea
one has \cite{cho1,cho2}
\bea
&&\delta A_\mu = \oneg \n \cdot \pro_\mu \valpha,\,\,\,\
\delta \hat A_\mu = \oneg \D_\mu \valpha  ,  \nn \\
&&\hspace{1.2cm}\delta \X_\mu = - \valpha \times \X_\mu  .
\eea
This shows that $\hat A_\mu$ by itself describes an $SU(2)$ connection which
enjoys the full $SU(2)$ gauge degrees of freedom. Furthermore 
$\vec X_\mu$ transforms covariantly under the gauge transformation. 
This confirms that our decomposition provides a gauge-independent 
decomposition of the non-Abelian potential into the restricted part 
$\hat A_\mu$ and gauge covariant part $\vec X_\mu$. 
This decomposition, which has recently
become known as the ``Cho decomposition'' \cite{faddeev2}
or the ``Cho-Faddeev-Niemi-Shabanov 
decomposition'' \cite{gies}, was
introduced long time ago in an attempt to demonstrate 
the monopole condensation in QCD \cite{cho1,cho2}. 
But only recently the importance of the decomposition 
in clarifying the non-Abelian dynamics 
has become appreciated by many authors \cite{faddeev2,lang}.
Indeed it is this decomposition which has played a crucial role to establish 
the possible connection between the Skyrme-Faddeev action and the
effective action of QCD \cite{cho3,lang,gies}, and the Abelian dominance
in the Wilson loops in QCD \cite{cho4}. 

To understand the physical meaning of our decomposition notice that  
the restricted potential $\hat{A}_\mu$ actually has a dual
structure.
Indeed the field strength made of the restricted potential is decomposed as
\begin{eqnarray}
& \hat{F}_{\mu\nu} = (F_{\mu\nu}+ H_{\mu\nu})\hat{n}\mbox{,}\nonumber \\
& F_{\mu\nu} = \partial_\mu A_{\nu}-\partial_{\nu}A_\mu \mbox{,}\nonumber \\
& H_{\mu\nu} = -\dfrac{1}{g} \hat{n}\cdot(\partial_\mu
\hat{n}\times\partial_\nu\hat{n})
= \partial_\mu \tilde C_\nu-\partial_\nu \tilde C_\mu,
\end{eqnarray}
where $\tilde C_\mu$ is the ``magnetic'' potential
\cite{cho1,cho2}. So one can identify the non-Abelian
monopole potential by
\bea
\vec C_\mu= -\frac{1}{g}\hat n \times \partial_\mu\hat n ,
\eea
in terms of which the magnetic field is expressed by
\bea
\vec H_{\mu\nu}&=&\partial_\mu \vec C_\nu-\partial_\nu \vec C_\mu+ g \vec
C_\mu \times \vec C_\nu
=-\frac{1}{g} \partial_\mu\hat{n}\times\partial_\nu\hat{n} \nn\\
&=&H_{\mu\nu}\hat n.
\eea
Notice that the magnetic field has a remarkable structure
\bea
H_{\mu\alpha}H_{\alpha\beta}H_{\beta\nu} =-\dfrac{1}{2}
H^2_{\alpha\beta}H_{\mu\nu},
\eea
which will be useful for us in the following.

Another important feature of $\hat{A}_\mu$ is that, 
as an $SU(2)$ potential, it retains the full 
topological characteristics of the original non-Abelian potential.
Clearly the isolated singularities of $\hat{n}$ defines $\pi_2(S^2)$
which describes the non-Abelian monopoles.  Indeed $\hat A_\mu$
with $A_\mu =0$ and $\hat n= \hat r$ describes precisely
the Wu-Yang monopole \cite{wu,cho5}.  Besides, with the $S^3$
compactification of $R^3$, $\hat{n}$ characterizes the
Hopf invariant $\pi_3(S^2)\simeq\pi_3(S^3)$ which describes 
the topologically distinct vacuua \cite{bpst,thooft}.
This tells that the restricted gauge theory made of $\hat A_\mu$ 
could describe the dual dynamics which should play an essential
role in $SU(2)$ QCD \cite{cho1,cho2}.

With our decomposition (1) we have
\bea
\vec{F}_{\mu\nu}&=&\hat F_{\mu \nu} + \D _\mu \X_\nu -
\D_\nu \X_\mu + g\X_\mu \times \X_\nu,
\eea
so that the Yang-Mills Lagrangian is expressed as
\bea
{\cal L} = &-&\dfrac{1}{4} \vec F^2_{\mu \nu }
=-\dfrac{1}{4}
{\hat F}_{\mu\nu}^2 -\dfrac{g}{2} {\hat F}_{\mu\nu} \cdot (\X_\mu \times \X_\nu)  \nn \\
 &-&\dfrac{1}{4}(\D_\mu\X_\nu-\D_\nu\X_\mu)^2-\dfrac{g^2}{4} (\X_\mu \times \X_\nu)^2. 
\eea
This shows that the Yang-Mills theory can be viewed as
the restricted gauge theory made of the Abelian projection,
which has an additional gauge covariant charged vector field
(the valence gluons) as its source \cite{cho1,cho2}. 

With this preparation
we will now derive the effective action of QCD 
in the presence of the monopole background, using the background field method
\cite{dewitt,pesk}. So we first devide
the gauge potential $\vec A_\mu$ into two parts, the slow-varying
classical part $\vec A^{(c)}_\mu$ and the fluctuating quantum
part $\vec A^{(q)}_\mu$, and identify the monopole potential $\vec C_\mu$
as the classical background,
\bea
&\vec A_\mu = \vec A^{(c)}_\mu + \vec A^{(q)}_\mu, \nn\\
&\vec A^{(c)}_\mu = \vec C_\mu,~~~~~\vec A^{(q)}_\mu = A_\mu \hat n + \X_\mu.
\eea
With this we introduce two types of gauge transformations, the background
gauge transformation and the physical gauge transformation.
Naturally we identify the background gauge transformation as
\bea
&\delta \vec C_\mu = \dfrac{1}{g} \bar D_\mu \vec \alpha,\nn\\
&\delta (A_\mu \hat n + \X_\mu) = - \vec \alpha \times
(A_\mu \hat n + \X_\mu),
\eea
where now $\bar D_\mu$ is defined with only
the background potential $\vec C_\mu$
\bea
\bar D_\mu = (\partial_\mu + g \vec C_\mu \times).
\eea
As for the physical gauge transformation which leaves the
background potential invariant, we must have
\bea
\delta \vec C_\mu = 0,~~~~~\delta (A_\mu \hat n + \X_\mu) =
\dfrac{1}{g} D_\mu \vec \alpha.
\eea
Notice that both (12) and (14) respect the original
gauge transformation,
\bea
\delta \A_\mu =
\dfrac{1}{g} D_\mu \vec \alpha. \nonumber
\eea
Now, we fix the gauge by
imposing the following gauge condition to the quantum fields,
\bea
&{\vec F}=\bar D_\mu (A_\mu \hat n + \vec X_\mu) =0 , \nn\\
&{\cal L}_{gf} =- \dfrac{1}{2\xi}
\left[(\partial_\mu A_\mu)^2 + ({\bar D}_\mu \X_\mu)^2\right].
\eea
The corresponding Faddeev-Popov determinant is given by
\bea
M^{ab}_{FP} = \dfrac {\delta F^a}{\delta \alpha^b} = (\bar D_\mu D_\mu)^{ab}.
\eea
With this gauge fixing
the effective action takes the following form,
\bea
\Gamma (\vec C_\mu)&=&\int {\cal D} A_\mu {\cal D}
\X_\mu {\cal D} \vec{c} {\cal D}\vec{c}^{~*}
\exp \{{~i \int[-\dfrac {1}{4}{\hat F}_{\mu \nu}^2} \nn \\
&-&\dfrac{1}{4} ( \D_\mu \X_\nu -\D_\nu \X_\mu)^2
-\dfrac{g}{2} {\hat F}_{\mu\nu} \cdot (\X_\mu \times \X_\nu) \nn\\
&-&\dfrac{g^2}{4}(\X_\mu \times \X_\nu)^2
+\vec{c}^{~*}\bar {D}_\mu D_\mu\vec{c} \nn\\
&-&\frac{1}{2\xi}(\partial_\mu A_\mu)^2-\frac{1}{2\xi}
(\bar {D}_\mu\vec{X}_\mu)^2]d^4x\},
\eea
where $\vec c$ and ${\vec c}^{~*}$ are the ghost fields.
Notice that the effectice action is explicitly invariant
under the background gauge transformation (12),
if we treat the ghost fields as quantum fields and add the following
gauge transformation of the ghost fields to (12),
\bea
\delta \vec c = - \alpha \times \vec c,
~~~~~\delta \vec c^{~*} = - \alpha \times \vec c^{~*}.
\eea
This guarantees that the resulting effective action we obtain
after the functional integral should be invariant under
the remaining background gauge transformation which involves
only $\vec C_\mu$. This, of course, is the advantage of the
background field method which greatly simplifies the calculation
of the effective action \cite{dewitt,pesk}.

Now, we can perform the functional integral in (17). Remember that in one loop
approximation only the terms quadratic in quantum fields
become relevant in the functional integral. 
So the $A_\mu$ integration becomes trivial,
and the $\X_\mu$ and ghost integrations result in the
following functional determinants (with $\xi=1$),
\bea
&{\rm Det}^{-\frac{1}{2}} K_{\mu \nu}^{ab}\simeq
{\rm Det}^{-\frac{1}{2}}[g_{\mu \nu}
 (\bar D \bar D)^{ab}
- 2gH_{\mu \nu}\epsilon^{abc} n^c], \nn\\
&{\rm Det} M^{ab}_{FP} \simeq {\rm Det} (\bar{D} \bar{D})^{ab}.
\eea
One can simplify the determinant $K$ 
using the relation (8),
\bea
\ln {\rm Det}^{-\frac{1}{2}} K& =&-\ln {\rm Det}(\bar{D}\bar{D})^{ab}\nn\\
&-&\frac12\ln {\rm Det}[(\bar{D}\bar{D})^{ab}
+i\sqrt{2}gH\epsilon^{abc}n^c]\nn\\
&-&\frac12\ln {\rm Det}[(\bar{D}\bar{D})^{ab}-i\sqrt{2}gH\epsilon^{abc}n^c],
\eea
where $H=\sqrt{\vec{H}_{\mu\nu}^2}$.
With this the one loop contribution of the functional
determinants to the effective action can be written as
\bea
\Delta S &=& i\ln {\rm Det}(\tilde{D}^{2} +\sqrt{2}gH) \nn\\
         &+& i\ln {\rm Det} (\tilde{D}^{2} -\sqrt{2}gH),
\eea
where now $\tilde{D}_\mu$ acquires the following Abelian form,
\bea
\tilde{D}_\mu =\partial_\mu + ig\tilde{C}_\mu .\nn
\eea
Notice that the reason for this simplification is precisely because
$\vec{C}_\mu$ originates from the Abelian projection.
One can evaluate the functional determinants in (21) with the
Fock-Schwinger proper time method, and for a constant background 
$H$ we find
\bea
&\Delta{\cal L} = \dfrac{1}{16 \pi^2}\int_{0}^{\infty} 
\dfrac{dt}{t^2}
 \dfrac{g H/\sqrt{2} \mu^2}{\sinh (g H t/\sqrt{2}\mu^2)}\nn \\
&\times[ \exp (-\sqrt{2}g H t/\mu^2 )+  \exp (\sqrt{2}g H t/\mu^2)],
\eea
where $\mu$ is a dimensional parameter.
The integral contains the (usual) ultra-violet divergence 
around $t \simeq 0$, but notice that
it is also plagued by a severe infra-red divergence 
around $t \simeq \infty$. This is because 
the functional determinant (21) contains negative eigenvalues, 
whose eigenfunctions become
tachyonic in the infra-red region. More precisely when the momentum $k$
of the gluon parallel to the background magnetic field becomes smaller
than the background field strength (i.e., when $k^2 < gH/ \sqrt 2$),
the lowest Landau level gluon eigenfunction whose spin is parallel to
the magnetic field acquires an imaginary energy and thus becomes
tachyonic. So to calculate the functional determinant we must exclude these 
unphysical tachyonic modes which causes the infra-red instability.

The correct infra-red regularization is dictated by the causality.
To implement the causality in (22) we first go to the Minkowski time 
with the Wick rotation, and find
\bea
&\Delta{\cal L} = \Delta{\cal L_+} + \Delta{\cal L_-}, \nn\\
& \Delta{\cal L_+} =  - \dfrac{1}{16 \pi^2}\int_{0}^{\infty} 
\dfrac{dt}{t^2}
 \dfrac{gH/\sqrt{2} \mu^2}{\sin (gHt/\sqrt{2}\mu^2)}\nn \\
&\times[ \exp (-i\sqrt{2}g H t/\mu^2 )], \nn\\
& \Delta{\cal L_-} =  - \dfrac{1}{16 \pi^2}\int_{0}^{\infty} 
\dfrac{dt}{t^2}
 \dfrac{gH/\sqrt{2} \mu^2}{\sin (gHt/\sqrt{2}\mu^2)}\nn \\
&\times[ \exp (+i\sqrt{2}g H t/\mu^2 )]. 
\eea
In this form the infra-red divergence has disappeared,
but now we face the ambiguity in choosing the correct contours
of the integrals in (23). Fortunately this ambiguity can 
be resolved by the causality. Remember that the two integrals
$\Delta{\cal L_+}$ and $\Delta{\cal L_-}$ originate from the
two determinants in (21), and the standard causality argument requires us to
identify $\sqrt{2}g H$ in the first determinant as  
$\sqrt{2}g H -i\epsilon$ but in the second determinant as
$\sqrt{2}g H +i\epsilon$. This tells that
the poles in the first integral in (23) should lie above the real
axis, but the poles in the second integral should lie 
below the real axis. From this we conclude
that the contour in $\Delta{\cal L_+}$ should pass below the
real axis, but the contour in $\Delta{\cal L_-}$ should pass above the
real axis. With this causality requirement the two integrals 
become identical, so that we finally have
\bea
&\Delta{\cal L} = \dfrac{1}{16 \pi^2}\int_{0}^{\infty} 
\dfrac{ d t}{t^{2-\epsilon}}
 \dfrac{g  H/\sqrt{2} \mu^2  }{\sinh  (g H t/\sqrt{2}\mu^2 ) }\nn \\
&\times[ \exp (-\sqrt{2}g H t/\mu^2 )+  \exp (-\sqrt{2}g H t/\mu^2 )],
\eea
where now $\epsilon$ is the ultra-violet cutoff which we have introduced
to regularize the ultra-violet divergence. 

From these infra-red and ultra-violet regularization 
we finally obtain the following effective Lagrangian of QCD
(with the modified minimal subtraction)
\bea
&{\cal L}_{eff}=-\dfrac{1}{4}H^2 -\dfrac{11g^2}{96\pi^2}H^2(\ln
\dfrac{gH}{\mu^2}-c ), \nn\\
&c=1+\dfrac{15}{22}\ln2+\dfrac{12}{11}\zeta'(-1)=1.2921409..... ,
\eea
where now $\zeta(s)$ is the Riemann's zeta function.
This completes our derivation of the one-loop effective Lagrangian of $SU(2)$
QCD in the presence of the monopole background. Notice that, 
as expected, the effective lagrangian is explicitly invariant
under the background gauge transformation (12).

We emphasize that our infra-red regularization by causality naturally
excludes the tachyonic modes from the functional integral.
This is because the tachyonic modes will certainly violate
the causality, and should be forbidden by the causality. 
Including the tachyons 
in the physical spectrum will surely destablize
QCD and make it ill-defined.

Clearly the effective action provides the following non-trivial
effective potential
\bea
V=\frac14 H^2\Big[1+\frac{11 g^2}{24 \pi^2}(\ln\frac{gH}{\mu^2}-c)\Big],
\eea
which generates a non-trivial local minimum at 
\bea
<H>=\frac{\mu^2}{g} \exp\Big(-\frac{24\pi^2}{11g^2}+ c-\frac12\Big).
\eea
This is nothing but the desired magnetic condensation. 
{\it This proves that the one
loop effective action of QCD in the presence of the constant magnetic
background does generate a dynamical symmetry breaking thorugh the
monopole condensation}. 

To check the consistency of our result with the perturbative QCD
we now discuss the running coupling and the renormalization.
For this we define the running coupling $\bar g$ by
\bea
\frac{\partial^2V}{\partial H^2}\Big|_{H={\bar \mu}^2} 
=\frac{1}{2}\frac{g^2}{ \bar g^2}.
\eea
With this we obtain 
\bea
\frac{1}{\bar g^2} =
\frac{1}{g^2}+\frac{11}{24\pi^2}\Big(\ln\frac{g {\bar\mu}^2}{{\mu}^2}
-c + \frac{3}{2}\Big),
\eea
and the following $\beta$-function,
\bea
\beta(\bar\mu)=-\frac{11}{24\pi^2} \bar g^3~,
\eea
which exactly coincides with the well-known 
asymptotic freedom result \cite {gross}.
In terms of the running coupling the renormalized potential is given by
\bea
V_{\rm ren} \simeq \frac14 H^2\Big[1+\frac{11 \bar g^2}{24 \pi^2}
(\ln\frac{H}{\bar\mu^2}-\frac{3}{2})\Big].
\eea
From this we obtain the following Callan-Symanzik equation
\bea
\Big(\bar\mu\frac{\partial}{\partial \bar\mu}+\beta\frac{\partial}{\partial \bar g}
-\gamma\vec{C}_\mu\frac{\partial}{\partial\vec{C}_\mu} \Big)V_{\rm ren}\simeq 0,
\eea
where $\gamma$ is the anomalous dimension for $\vec C_\mu$,
\bea
\gamma=-\frac{11}{24\pi^2}\bar g^2.
\eea
This proves the renormalization group invariance of our effective potential. 

The fact that our effective potential (26) and 
the resulting vacuum (27) are expressed in terms of $H$ 
assures that they are explicitly gauge and Lorentz invariant. 
On the other hand notice that, in terms of the magnetic potential $\vec C_\mu$,
the effective potential could be written as
\bea
V&=&\frac{g^2}{4}(\vec{C}_\mu\times\vec{C}_\nu)^2\Big \{ 1 \nn \\
&+&\frac{11 g^2}{24 \pi^2}
\Big[ \ln \frac{g[(\vec{C}_\mu\times\vec{C}_\nu)^2]^{1/2}}
{\mu^2}-c\Big]\Big\},
\eea
Now, just for a heuristic reason, suppose we choose 
a particular Lorentz frame and 
express the vacuum (27) by the vacuum expectation value of $\vec C_\mu$.
In this picture the above effective potential generates the following
mass matrix for $\vec C_\mu$,
\bea
&{\displaystyle M}_{ij}^{\mu\nu} = \Big< \dfrac {\delta^2 V}
{\delta {C}^{i}_{\mu} \delta {C}^{j}_{\nu}} \Big> 
= m^2~(\delta_{ij}-n_i n_j) g^{\mu\nu}, \nn\\
&m^2=\dfrac{11g^4}{96\pi^2}\Big<\dfrac{(\vec{C}_\mu\times
\vec{H}_{\mu\nu})^2}{H^2}\Big>,
\eea
where $m$ can be interpreted as the ``effective mass'' for $\vec C_\mu$.
This demonstrates that the magnetic condensation indeed generates the
mass gap necessary for the dual Meissner effect and
the confinement.

With the above understanding we can now study the possible connection
between the Skyrme-Faddeev action and the effective actin of QCD.
To do this we first expand the effective potential in terms of
the magnetic potential around the vacuum and make the following
approximation,
\bea
V &\simeq& V_0 + \dfrac {1}{2!} {\Big <} \dfrac {\delta^2 V}{\delta C^i_\mu \delta
C^j_\nu}{\Big >}
~{\bar C}^i_\mu ~{\bar C}^j_\nu  \nn \\
  &+& \dfrac{1}{3!}  {\Big <} \dfrac {\delta^3 V}{\delta C^i_\mu \delta
C^j_\nu \delta C^k_\rho}{\Big >}
~{\bar C}^i_\mu ~{\bar C}^j_\nu ~{\bar
C}^k_\rho \nn \\
  &+& \dfrac{1}{4!}  {\Big <} \dfrac {\delta^4 V}{\delta C^i_\mu \delta
C^j_\nu \delta C^k_\rho \delta C^l_\sigma }{\Big >}
~{\bar C}^i_\mu ~{\bar C}^j_\nu ~{\bar C}^k_\rho ~{\bar C}^l_\sigma,
\eea
where 
\bea
{\bar C}^i_\mu = C^i_\mu - < C^i_\mu>. \nn
\eea
Notice that here we have neglected the higher order terms and 
kept only the quartic polynomial in $\vec C_\mu$
for simplicity. In this case the corresponding 
effective Lagrangian will acquire the form
\bea
{\cal L}_{eff}&\simeq&- \frac12m^2(\vec{C}_\mu) ^2 
- \frac{\alpha}{4} (\vec{C}_\mu\times\vec{C}_\nu)^2 \nn\\
&-& \frac{\beta}{4} (\vec C_\mu \cdot \vec C_\nu)^2 - \frac{\gamma}{4}
(\vec C_\mu)^4
+ ........\nn\\
&=&-\frac{m^2}{2g^2}(\partial_\mu \hat{n})^2
-\frac{\alpha}{4g^2}(\partial_\mu \hat{n}
\times\partial_\nu \hat{n})^2 \nn\\
&-& \frac{\beta}{4g^2} (\partial_\mu \hat n \cdot \partial_\nu \hat n)^2
- \frac{\gamma}{4g^2} (\partial_\mu \hat n)^4 + .......,
\eea
where $\alpha$, $\beta$, and $\gamma$ are numerical 
parameters determined by (36). 
This is nothing but a generalized Skyrme-Faddeev Lagrangian \cite {faddeev1}.
This shows that one can indeed derive a generalized Skyrme-Faddeev action
from QCD by expanding the effective potential around the vacuum. 

But there are two points that should be emphasized here. 
First, our approximation (37) is by no means exact. 
For example, we have assumed a constant magnetic background
to derive the effective action. 
So our analysis establishes a possible connection 
between a generalized non-linear sigma model of Skyrme-Faddeev type
and QCD only in a limited sense. In particular it does not assert
that the simple Skyrme-Faddeev action could describe QCD in the infra-red limit.
Secondly, we had to choose a particular Lorentz frame to justify
the expansion (36) of the effective action around the vacuum. 
So our argument appears to have compromised the Lorentz invariance,
although the final effective Lagrangian (37) is obviously Lorentz invariant.
In spite of these drawbacks our analysis strongly 
endorses the fact that the Skyrme-Faddeev action 
has something in common with QCD, which is truly remarkable.

Similar result has recently been obtained by 
many authors \cite{cho3,lang,gies}.
But a new feature in our analysis is that the resulting Skyrme-Faddeev action 
is intimately connected to the monopole condensation in QCD.
In particular our analysis makes it clear that the mass scale 
in the Skyrme-Faddeev action is
directly related to the mass of the monopole potential, 
which determines the confinement scale in QCD. This is not surprizing.
Indeed any attempt to relate the Skyrme-Faddeev action to QCD
must produce the mass scale that the Skyrme-Faddeev action contains,
and the only way to interpret this mass scale in QCD is through
the confinement.

It must be emphasized that our decomposition (1) has played
the central role in our calculation of
the effective action of QCD. It is this decomposition which allows
us to integrate out the gauge covariant part of the gluons without
breaking the gauge invariance, and reduce the non-Abelian dynamics 
effectively into an Abelian form which has a charged vector source.
In particular it is this decomposition which allows us 
the gauge-independent separation of the monopole background
from the dynamical degrees of the non-Abelian gauge theory.
We conclude with the following remarks: \\
1) One might question (legitimately)
the validity of the one loop approximation,
since in the infra-red limit the non-perturbative effect
is supposed to play the essential role
in QCD. Our attitude on this issue is that QCD can be viewed as the
perturbative extension of the topological field theory described
by the restricted QCD, so that the non-perturbative
effect in the low energy limit can effectively be represented by
the topological structure of the restricted gauge
theory. This is reasonable,
because the large scale structure of the monopole topology
naturally describes the long range behavior
of the theory. In fact one can show that it is the restricted connection
that plays a crucial role in the Wilson loop integral,
which provides the confinement criterion in QCD
\cite{cho4}.
So we believe that our classical monopole
background (with the monopole charge $1/g$) automatically
takes care of the essential feature of the non-perturbative effect,
which should make the one loop approximation reliable. \\
2) One may notice that our vacuum (27) looks very much like 
the old ``Savvidy vacuum'' \cite{savv}.
But, unlike the Savvidy vacuum, ours is stable. This
is precisely because our infra-red regularization by causality 
excludes the unphysical tachyonic modes
in our calculation of the functional integral (17). The
tachyonic modes in the functional integral, if included, will generate
an imaginary (absorptive) part in the effective action, which should
destablize the monopole condensation. Indeed this was the origin of the
instability of the Savvidy vacuum. We don't have this instability.
The stability of our vacuum is guaranteed by the causality. 
Of course one might question
the wisdom of our infra-red regularization by causality. 
Indeed the issue of what 
infra-red regularization procedure one should adopt is the central issue
in QCD, which certainly need a more careful examination.
Here we simply point out that 
we have other independent justifications which support
our infra-red regularization by causality \cite{sch,cho6}. Only with the
exclusion of the unphysical modes one can obtain a consistent 
theory of QCD. \\ 
3) We emphasize that, independent of the stability of the monopole
condensation, the real (dispersive) part of the effective action of QCD
can indeed be related to a generalized Skyrme-Faddeev action around the local
minimum. This is remarkable. On the other hand it should be emphasized 
that it is highly unlikely that the Faddeev-Niemi
knots can actually be realized in QCD. This is because the Faddeev-Niemi
knots are made of the (color) magnetic flux, while the QCD knots
(if possible at all) should consist of the (color) electric flux. So at best
the Faddeev-Niemi knots could be interpreted as a ``dual'' description 
of the possble QCD knots. If this is so, the really interesting question 
now is whether QCD could admit the color electric knots which
could be interpreted as the new glueball states. This question 
is worth a further investigation. 

It must be clear from our analysis that
the existence of the magnetic condensation (independent of its stability) 
is a generic
feature of the non-Abelian gauge theory. Whether this condensation
should describe the true vacuum of QCD or not, of course, is the most
important issue in QCD. This issue, and the
generalization of our result to $SU(3)$,
will be discussed in a forthcoming paper \cite{cho6}.

The work is supported in part by Korean Science and Engineering
Foundation (KRF-2000-015-BP0072), and by the BK21 project of 
Ministry of Education.

\end{document}